\documentclass[pra, twocolumn, article,nofootinbib, superscriptaddress]{revtex4-1}

\usepackage{amsmath,amssymb,amsthm}
\usepackage{quantum}
\usepackage{color}
\usepackage{tikz}
\usepackage{hyperref}

\begin{document}
\title{Linear-time general decoding algorithm for the surface code}

\author{Andrew S. Darmawan}
\affiliation{Department of Applied Physics, The University of Tokyo, 7-3-1 Hongo, Bunkyo-ku, Tokyo 113-8656, Japan}
\author{David Poulin}
\affiliation{D\'epartement de Physique and Institut Quantique, Universit\'e de Sherbrooke, Qu\'ebec, Canada J1K 2R1}
\affiliation{Canadian Institute for Advanced Research, Toronto, Ontario, Canada M5G 1Z8}

\begin{abstract}
A quantum error correcting protocol can be substantially improved by taking into account features of the physical noise process. We present an efficient decoder for the surface code which can account for general noise features, including coherences and correlations. We demonstrate that the decoder significantly outperforms the conventional matching algorithm on a variety of noise models, including non-Pauli noise and spatially correlated noise. The algorithm is based on an approximate calculation of the logical channel using a tensor-network description of the noisy state.
\end{abstract}

\date{\today}

\maketitle

\noindent{\em Introduction. --- }
The surface code represents a promising route toward universal quantum computing. It achieves high performance with a simple, two-dimensional physical layout and it is therefore the focus of many current experimental efforts \cite{kelly_state_2015, takita_demonstration_2016}. While the emphasis is often placed on making greater distance surface codes with more physical qubits, in this Rapid Communication we demonstrate that the error-correcting power of the surface code may be substantially improved merely by upgrading the classical control software, without expending any additional hardware resources. 

The control software operates a {\em decoding algorithm}, which selects a correction given the error syndrome. The original decoder for the surface code is minimum-weight perfect matching (MWPM), which selects the correction with the smallest weight compatible with the syndrome. \cite{dennis_topological_2002} 
While MWPM's simplicity is appealing, it does not consider the underlying noise process and, as a result, will generally perform suboptimally. 

Some simple noise properties ignored by matching include error degeneracy (i.e. the fact that distinct errors have the same effect on the code) and correlations between single-qubit errors in conjugate basis (Pauli $X$ and $Z$ errors). Improved decoders have been devised which account for these properties. For instance, \cite{bravyi_efficient_2014-1} accounts for degeneracy with an exact decoder built on matchgate circuits and $X$-$Z$ correlations using an approximate mapping to a one-dimensional quantum chain. 

Real noise has many additional features which may be used to further improve decoding. For instance, using inefficient brute-force calculations, it has been shown that a non-negligible improvement can be obtained over MWPM by accounting for spatial noise correlations \cite{heim_optimal_2016}. Even when restricted to local noise, other noise features can greatly impact a code's performance \cite{chamberland_hard_2017,iyer_small_2017}. 

A number of other decoders have been developed with advantages over MWPM in terms of speed or accuracy \cite{ harrington_analysis_2004, duclos-cianci_fast_2010, wootton_high_2012, bravyi_quantum_2013, duclos-cianci_fault-tolerant_2014, hutter_efficient_2014, watson_fast_2015,herold_cellular-automaton_2015, herold_cellular_2017, torlai_neural_2017, baireuther_machine-learning-assisted_2018, krastanov_deep_2017-1, delfosse_almost-linear_2017, tuckett_ultrahigh_2018}. However, the noise processes considered in those works were always assumed to be uncorrelated Pauli  noise, with the exception of \cite{nickerson_analysing_2017} which is tailored to a specific form of correlated Pauli noise.

In this Rapid Communication, we present a decoding algorithm for the surface code which can account for general noise, including non-Pauli noise and spatially correlated noise. 
A general noise model is specified by an $N$-qubit  completely positive trace-preserving (CPTP) map $\mathcal N$. 
Our decoder is tailored to any CPTP map representable by a two-dimensional tensor network, which includes arbitrary local noise, as well as quite general spatially correlated noise. In essence, such noise models describe any noise process where spatial correlations are mediated by short-range interactions directly between qubits or indirectly through localized environmental degrees of freedom.  

Our decoder performs an approximate calculation of the logical channel that has affected the encoded data, and chooses the correction which best inverts it so that the overall action on the encoded data is as close as possible to the identity.  
This calculation relies on the projected entangled pair operator (PEPO) description of the noisy state, which is guaranteed by the above assumption about the noise. 

We apply our decoder to examples of non-Pauli and spatially correlated noise and observe orders of magnitude improvement over MWPM. For the local noise models tested the decoder appears to perform near optimally. The cost of the decoder is $O(ND^3\chi^3)$, where $N$ is the number of physical qubits, $D$ is the bond dimension of the noise CPTP map ($D=1$ for uncorrelated noise), and $\chi$ is the bond dimension used in the approximate tensor network contraction. In the noise models studied, a small constant value of $\chi$ of about $8$ appears sufficient to substantially outperform matching.

Our paper is structured as follows. We first introduce basic concepts and outline the decoding algorithm. We then present numerical results, and discuss future research directions. 

\medskip

\begin{figure}
    \includegraphics[width=0.4\textwidth]{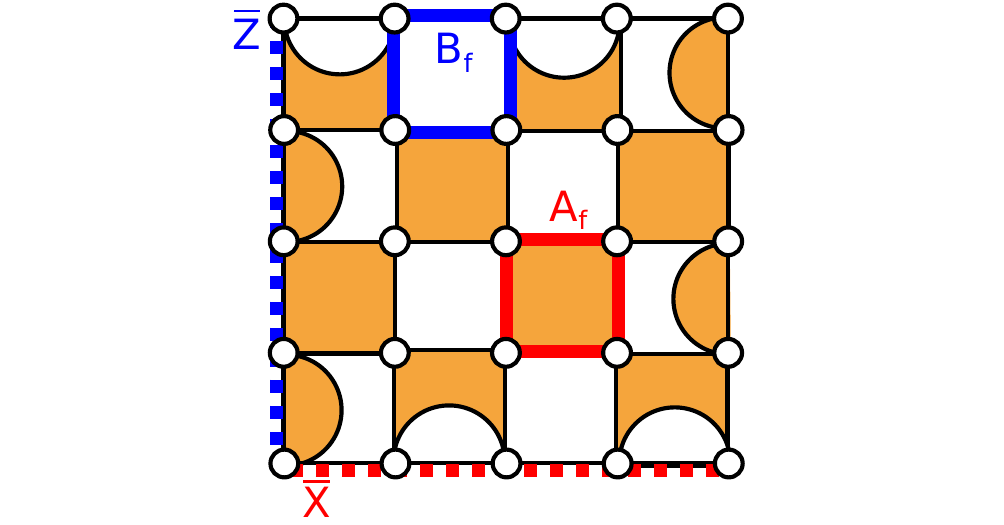}
    \caption{Layout of the surface code where qubits are located at vertices. Orange faces represent $A_f$ checks and white faces represent $B_f$ checks. The logical $\overline{Z}$ operator is a product of $Z$ along the dashed blue and the logical $\overline{X}$ operator is a product of $X$ along the dashed red line.}
    \label{f:surface_code}
\end{figure}

\noindent{\em Surface code. --- } The surface code, for our purposes, consists of a square lattice of $N$ qubits with open boundary conditions on which a set of commuting check operators is defined.  The layout of check operators follows that of \cite{bombin_optimal_2007} and is illustrated in Fig. \ref{f:surface_code} . On every white face $f$, an $x$-check operator is defined as $A_f = \prod_{i\in f} X_i$, where the product is taken over all vertices surrounding the face. Likewise, on every orange face $f$ a $z$-check operator is defined as $B_f = \prod_{i\in f} Z_i$. There are $N-1$ checks, all of which commute, and the group generated by all check operators is called the stabilizer of the code. The codespace is defined as the simultaneous $+1$ eigenspace of all check operators. By definition, any operator in the stabilizer acts as the identity operator on the codespace. With the specified check layout, the codespace is two dimensional, in other words, the surface code encodes a single qubit. The logical $\overline{Z}$ operator is given as a string of Pauli $Z$ operators along the left boundary of the code. This operator commutes with the every check, and therefore preserves the codespace. However, it is not contained in the stabilizer of the code, and its action on the codespace is not the identity. The encoded qubit states $\ket{0}_L$ and $\ket{1}_L$ can be defined as the $+1$ and $-1$ eigenstates, respectively, of $\overline{Z}$ in the codespace. Therefore $\overline{Z}$ acts as Pauli $Z$ on the encoded qubit. Likewise, a logical $\overline{X}$ operator, which acts as Pauli $X$ on the encoded qubit, is given as a product of Pauli $X$ operators along the bottom boundary of the code. 

During a round of error correction, every check is measured. When physical qubits are affected by noise, $-1$ measurement outcomes will occur with nonzero probability and the set of measurement outcomes $s$ is called the syndrome. Using the syndrome one then applies further operations to the code qubits to return the state to the codespace, and undo any undesired transformation to the encoded information that may have occurred. The decoding problem is to determine the best correction operation to apply, based on this syndrome and knowledge of the noise model. We make this more precise in the following section. 

\medskip

\noindent{\em Decoding. --- } Here we define the decoding problem, which we express in a general form that is not restricted to stochastic Pauli noise. For concreteness, we decompose the physical evolution of the system during error correction into three distinct operators: noise $\mathcal{N}$, recovery $\mathcal{R}_s$, and decoder correction $\mathcal{D}_{s,\mathcal{N}}$. The latter two stages are dependent on the observed syndrome $s$, and the decoder correction can also depend on the noise model. 

The noise operation consists of application of the CPTP map $\mathcal{N}$ to the physical qubits of the surface code. After noise application, the code is no longer in the codespace. 

The recovery map is then the operation that returns the noisy state to the codespace. This consists of the syndrome measurement, which yields syndrome $s$ with probability $p(s)$ and which projects the state into an eigenspace of every check operator, followed by a Pauli operator that returns the state to the $+1$ eigenspace of every check operator. Note that this recovery simply returns the state to the codespace, without attempting to minimize the probability of a logical error. Given a syndrome $s$, we take this Pauli operator to be a product of strings of $Z$ operators that connect every flipped $x$ check to the top boundary and a product of strings of $X$ operators that connect every flipped $z$ check to the left boundary. 
The combined effect of the noise and recovery $\mathcal{E}_{s,\mathcal{N}}:=\mathcal{R}_s\circ \mathcal{N}$ is thus a map from the codespace to itself, so it is a single-qubit CPTP map. However, depending on the measured syndrome and the noise model, this $\mathcal{E}_{s,\mathcal{N}}$ may act on the encoded qubit in a nontrivial way. The goal of error correction is to preserve the state of the logical qubit, i.e., to make the overall transformation of the logical qubit as close to the identity as possible. For this, the final decoding step is necessary. 

In the final decoding step, a classical algorithm takes the measured syndrome and some description of the noise process and outputs a logical Pauli operator  $\mathcal{D}_{s,\mathcal{N}} \in\{\overline{I}, \overline{X}, \overline{Y}, \overline{Z}\}$. The target output of the decoding algorithm is the $\mathcal{D}_{s,\mathcal{N}}$ that best inverts $\mathcal{E}_{s,\mathcal{N}}$, i.e., minimises the logical error $\epsilon=||\mathcal{D}_{s,\mathcal{N}}\circ \mathcal{E}_{s,\mathcal{N}} - I||$. 
The norm defining $\epsilon$ can be taken to be any operator norm.

The computed decoder correction can be thought of as being applied to the code after recovery, however we remark that the Pauli operators involved in the recovery and decoding steps need not actually be applied in practice: keeping track of them is sufficient. 

The problem of finding a $\mathcal{D}_{s,\mathcal{N}}$ given $s$ and $\mathcal{N}$ that exactly minimizes the error appears hard. With the exception of some specific noise models, most decoders will only yield an approximation, so that for a non-zero fraction of the syndromes $s$, the selected $\mathcal{D}_{s, \mathcal{N}}$ does not minimize $||\mathcal{D}_{s, \mathcal{N}} \circ \mathcal{E}_{s, \mathcal{N}} - I||$ . We can quantify the performance of the decoder by the averaged logical error rate over all possible syndromes $\sum_s p(s)||\mathcal{D}_{s, \mathcal{N}} \circ \mathcal{E}_{s, \mathcal{N}} - I||$. 



%
%

\medskip 

\noindent{\em Decoding algorithm. --- } Here we describe our surface-code decoding algorithm. The essential idea is to compute an approximation $\tilde{\mathcal{E}}_{s, \mathcal{N}}$ of the logical channel $\mathcal{E}_{s, \mathcal{N}}$ given the syndrome $s$ and a CPTP map $\mathcal{N}$ of the noise. Once $\tilde{\mathcal{E}}_{s, \mathcal{N}}$ is known, the approximate logical error rate $\tilde{\epsilon}:=||L\circ \tilde{\mathcal{E}}_{s,\mathcal{N}} - I||$ is calculated for each $L\in \{\overline{I}, \overline{X}, \overline{Y}, \overline{Z}\}$. The decoder correction $\mathcal{D}_{s,\mathcal{N}}$ is then chosen to be the Pauli operator that minimizes $\tilde{\epsilon}$.\footnote{Note that in principle, this decoding algorithm could select a correction among all allowed logical operations of the code \cite{chamberland_hard_2017, chamberland_fault-tolerant_2018}; however, we restrict to Pauli corrections to be on the same footing as MWPM.} 

The nontrivial part of the algorithm is calculating a sufficiently accurate approximation of the logical channel $\tilde{\mathcal{E}}_{s,\mathcal{N}}$. For this, we draw on a simulation algorithm that we developed in \cite{darmawan_tensor-network_2017}. The key idea is that $\mathcal{E}_{s,\mathcal{N}}$ can be exactly expressed as a square-lattice tensor network. 

We start with a PEPO description of the codespace projector $\Pi_C$ and from this we compute a PEPO description of the noisy state $\mathcal{N}(\Pi_C)$\footnote{Note that in \cite{darmawan_tensor-network_2017} we started with a PEPO description of a half-encoded Bell state, rather than the code projector. However, these two descriptions are ultimately equivalent.}. This is possible if the noise $\mathcal{N}$ is an operator that maps PEPOs to PEPOs, so naturally any noise model $\mathcal N$ that can itself be represented by a two-dimensional tensor network will do. This is the only assumption we make about the noise. 
The recovery map, consisting of check measurements and a Pauli operator, can also be represented by a two-dimensional tensor network because it is built from a finite-depth local circuit. Thus, the output  $\mathcal{E}_{s,\mathcal{N}}(\Pi_0)=\mathcal{R}_s\circ\mathcal{N}(\Pi_0)$ of the recovery is also a PEPO, which is simple to calculate.

The map $\mathcal{E}_{s,\mathcal{N}}$ if fully characterized by its corresponding Choi matrix $C_{ij}={\rm tr}[L_i{\mathcal E}_{s,\mathcal{N}}(L_j\Pi_C)]$ describing the action of the channel on the Pauli basis, where $L_i= (\overline{I}, \overline{X}, \overline{Y}, \overline{Z})$ are logical Pauli operators. The operator $L_i{\mathcal{E}}_{s,\mathcal{N}}(L_j)$ is representable as a PEPO, and can be obtained for each $i,j$ by inserting tensors appropriately into the PEPO describing $\mathcal{E}_{s,\mathcal{N}}(\Pi_C)$. Full details of how these tensors are inserted are provided in \cite{darmawan_tensor-network_2017} and accompanying material. Given $L_i{\mathcal E}_{s,\mathcal{N}}(L_j)$, the target $C_{ij}$ can then be calculated by taking the trace of each pair of physical indices then contracting all indices in the resulting square-lattice tensor network.

Contracting a square-lattice tensor network is, in general, \#P-complete \cite{schuch_computational_2007}, and therefore no efficient algorithm is believed to exist. However, several algorithms exist that can output an approximate contraction in polynomial time \cite{verstraete_renormalization_2004, levin_tensor_2007, gu_tensor-entanglement-filtering_2009, lubasch_unifying_2014, evenbly_tensor_2015-1, yang_loop_2017}. We use an approximate algorithm for the contraction which involves treating the left-hand boundary of the tensor network as a one-dimensional spin chain, and approximately tracking its evolution as columns are applied to it \cite{schollwock_density-matrix_2011}. This contraction algorithm was also used in the context of surface-code error correction in other work \cite{bravyi_efficient_2014-1, darmawan_tensor-network_2017}. It has a time complexity of $O(N\chi^3)$ where $\chi$ is the bond dimension that controls the accuracy of the approximate contraction. Exact contraction requires that $\chi$ grows exponentially with $N$; however, as we will show, we have found that fixing $\chi$ to a small constant yields very accurate results.



\medskip

\noindent{\em Noise models. --- } We now describe the noise models that we have used to benchmark our decoder. We focus on two models to highlight noise features not considered in previous work. One feature is our decoder's ability to fully incorporate non-Pauli noise. To illustrate the performance of the decoder on non-Pauli noise we consider the amplitude-damping (AD) channel
$\mathcal{E}_{\rm AD}(\rho)=\sum_i K_i \rho K_i^\dag$, which has two Kraus operators,
\begin{equation}
    K_0 = \ketbra{0}{0} + \sqrt{1-\gamma}\ketbra{1}{1}\,,\quad K_1 = \sqrt{\gamma}\ketbra{0}{1}\,,
\end{equation}
where $\gamma\in [0,1]$ is the damping parameter. The full noise model is the $N$-fold tensor product of this single-qubit amplitude damping channel $\mathcal{N}_{\rm AD} = \mathcal{E}_{\rm AD}^{\otimes N}$.

Another feature of our decoder is that it can incorporate quite general spatial noise correlations. To demonstrate the performance of the decoder on spatially correlated noise, we define a correlated bit-flip noise model as follows. 

We specify an error on the surface code with a configuration of $N$ binary variables $\sigma=\sigma_1\sigma_2\cdots\sigma_N$ where $\sigma_i=1$ means that the qubit $i$ is unaffected while $\sigma_i=-1$ means that qubit $i$ is flipped. For this noise model, an error $\sigma$ occurs on the surface code with Boltzmann probability $p(\sigma) = e^{-\beta E(\sigma)}/ \mathcal{Z} $, where $\mathcal{Z}=\sum_{\sigma'} e^{-\beta E(\sigma')}$ is the partition function, $\beta$ is the inverse temperature parameter, and the energy $E(\sigma)$ is given by 
\begin{equation}
E(\sigma)=-h\sum_{i} \sigma_i - J_1 \sum_{\langle i,j \rangle} \sigma_i \sigma_j-J_2 \sum_{f} \left(\prod_{i \in f }\sigma_i \right)\,.
\end{equation}
The first sum is over all sites, the second sum is over all pairs of neighboring sites, and the final sum is taken over all white faces. The probability distribution $p(\sigma)$ therefore depends on three parameters $h$, $J_1$, and $J_2$ and the inverse temperature $\beta$ is an overall scaling parameter. The $h$ parameter influences the number of bit flips, with larger positive $h$ favoring configurations with fewer flipped spins. The $J_1$ parameter influences correlations between neighboring bit flips: if $J_1>0$ then the probability that a bit flip will occur at a site increases with the number of neighbours that are also flipped. Finally, the $J_2$ parameter influences the number of checks with $-1$ outcomes: larger $J_2$ favors configurations corresponding to syndromes with fewer flipped checks. Equivalently, increasing $J_2$ decreases the number of detectable error strings.  

This correlated bit-flip (CBF) noise model can be expressed as the nonlocal CPTP map 
\begin{equation}
\mathcal{N}_{\rm CBF}(\rho)=\sum_\sigma p(\sigma)U(\sigma)\rho U(\sigma)^\dag\,,
\end{equation}
where $U(\sigma)$ is the unitary map that applies an $X$ bit flip to sites $i$ with $\sigma_i=-1$, and the identity to every other site. Because any Boltzmann distribution of a local Hamiltonian is a local tensor-network \cite{hastings_inference_2008}, it is straightforward to show that $\mathcal{N}_{\rm CBF}$ can be expressed as a two-dimensional tensor network. 

We remark that, while the $J_2$ term affects $p(\sigma)$, it does not affect decoding since, once the syndrome $s$ is fixed, the probability of any error $\sigma$ consistent with the syndrome $s$ is independent of $J_2$. Hence, the noise model input to the decoder $\mathcal{N}_{\rm CBF}$ can be replaced with one with $J_2=0$, which simplifies the corresponding tensor network. The bond dimension of the noise tensor network is $D=2$. 


\medskip
\noindent{\em Results. --- } We tested our decoder on the above noise models at both high and low noise rates. While our decoding algorithm is efficient and can handle large lattices, verifying the quality of the decoder requires a full quantum-mechanical simulation. Amplitude damping and correlated bit-flip noise cannot be simulated in the same way and therefore require different benchmarking methods. 

In general, it is not possible to efficiently simulate non-Pauli noise, like amplitude damping. Therefore we have performed simulations on small system sizes where simulation is possible. We have used the exact (albeit inefficient) simulation algorithm described in \cite{darmawan_tensor-network_2017} which is essentially the same as the present decoding algorithm except that the logical channel is calculated exactly using an exact contraction of the tensor network, rather than an approximate contraction. With this algorithm, benchmark simulations could be performed on system sizes of up to $9\times 17$. Since the simulation algorithm outputs the exact logical channel, it is possible to compare our decoder with an optimal decoder, in which the correction is chosen using the exact logical channel, rather than an approximate one. 

For correlated bit-flip noise, it is possible to efficiently simulate the noise by sampling errors $\sigma$ from the probability distribution $p(\sigma) = e^{-\beta E(\sigma)}/ \mathcal{Z} $. Sampling from the distribution $p(\sigma)$ can be done efficiently using standard Markov chain Monte Carlo algorithms. For any sampled error, the syndrome is unambiguous and the correction can be calculated  using the decoding algorithm. The logical error is determined by calculating the resulting homology of the error combined with the recovery and decoding correction. If the resulting homology is trivial no logical error occurred, otherwise an undesired logical $\overline X$ error was applied to the encoded qubit.  

\begin{figure}[t]
    \centering
    \includegraphics[width=0.35\textwidth]{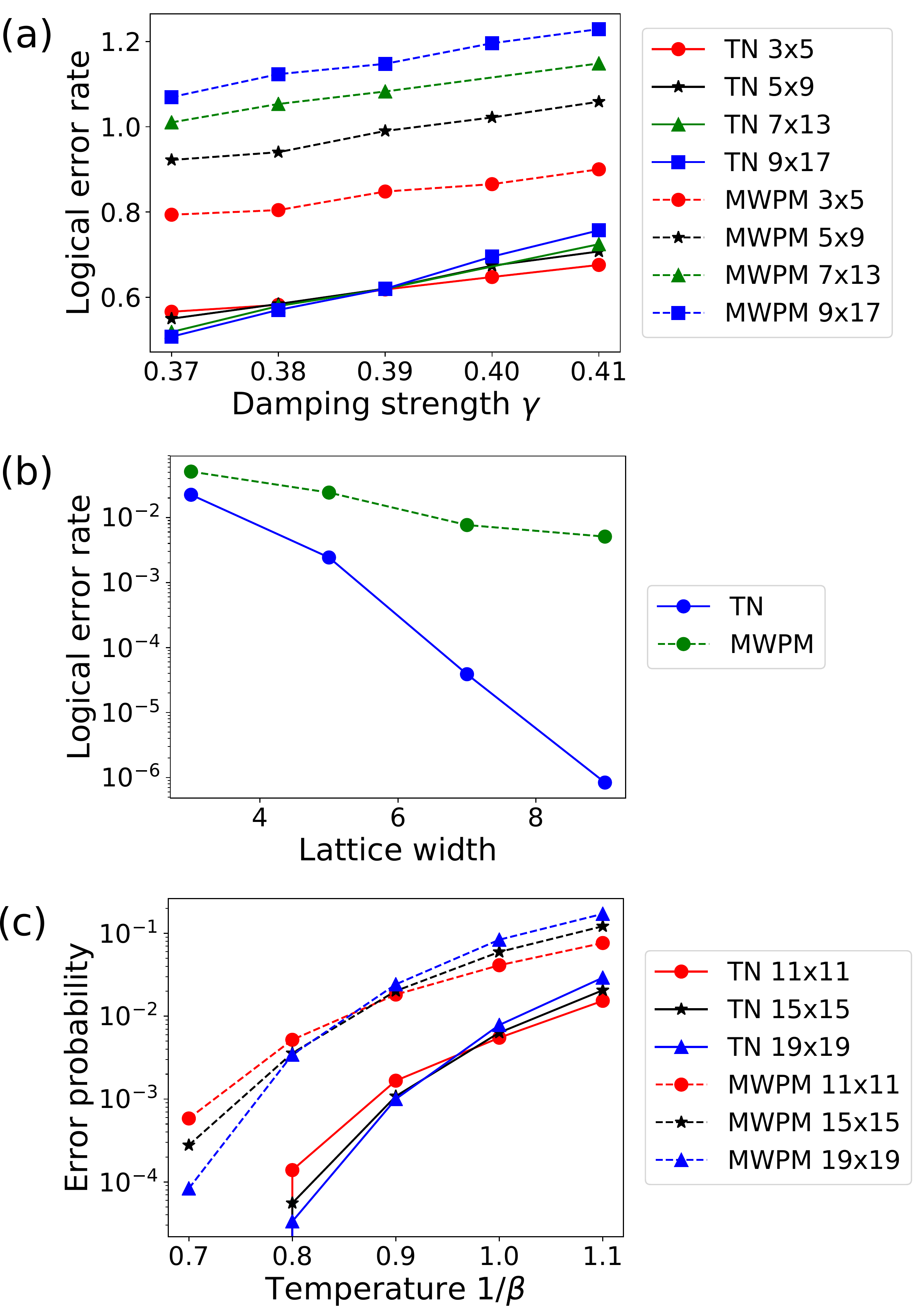}

    \caption{Decoder performance under high-strength amplitude damping (a), low-strength amplitude damping (b) and correlated bit-flip noise (c). For amplitude damping, the logical error rate is expressed in terms of the diamond distance of the logical channel from the identity, which we have computed using our surface-code simulation algorithm \cite{darmawan_tensor-network_2017}. In (a),  error rates are close to optimal threshold ($\gamma=39\pm1\%$) \cite{darmawan_tensor-network_2017}, and $\gamma$ is varied.   In (b), a low error rate $\gamma=9\%$ is fixed, and the lattice width $W$ is varied. In both cases, an asymmetric lattice with length $2W-1$ (i.e., longer logical $\overline X$) was used because amplitude damping has a greater tendency to flip $z$ checks than $x$ checks. For correlated bit-flip noise (c), the $y$-axis is the probability of a logical error, which is estimated by sampling over physical error configurations. For the TN decoder, no logical errors were observed for $1/\beta\le0.7$, implying logical error rates of less than $3\times 10^{-5}$.}
    \label{f:decoder_performance}
\end{figure}

In all of the following calculations the accuracy in the approximate algorithm $\chi=8$ was fixed. Between $1.2\times10^4$ and $1.2\times 10^5$ samples were taken for each data point. In Figs. \ref{f:decoder_performance}(a) and \ref{f:decoder_performance}(b) we present results for high- and low-strength amplitude damping. In the low-strength case, the performance of our tensor-network (TN) decoder was indistinguishable from that of the optimal decoder, and corresponds to an improvement of several orders of magnitude over MWPM. In the high noise case, for the largest system size tested, the difference in error rate (as measured by the diamond distance from the identity) between the TN decoder and optimal decoder was less than $0.01$, compared to around $0.6$ for MWPM.


In Fig. \ref{f:decoder_performance}(c) we show the results for correlated bit-flip noise. We fixed $J_1=1$, $J_2=-1.5$, $B=0.01$ and controlled the noise strength by varying the inverse temperature $\beta$. Noise strength was varied from high (above an apparent threshold) to low (where logical errors became undetectable). We observed a substantial improvement over matching for the entire range of noise strengths.

\medskip
\noindent{\em Conclusion. --- }We have presented a new decoder for the surface code which achieves high performance by exploiting information about the noise process. The decoder can account for a wide variety of noise properties, including spatial correlations and coherences. We have tested the decoder on spatially correlated bit-flip noise and amplitude damping, and have observed a large improvement over MWPM. 

Since our algorithm essentially maps the general decoding problem to the problem of contracting a tensor network, a wide range of tensor-network techniques may be employed to further improve the algorithm. For instance, the tensor contraction could be parallelized using a renormalization strategy to contract the network \cite{evenbly_tensor_2015-1, yang_loop_2017}. This would reduce the runtime scaling in $N$ from $O(N)$ to $O(\log N)$. Furthermore, generalization of our decoder to other local stabilizer codes is straightforward and depends only on whether there exists an efficient contraction algorithm for the tensor network representing the code's codespace. 

We have assumed that syndrome measurements are performed noiselessly. To generalize the decoding algorithm to imperfect measurements, the syndrome needs to be observed over time. For our decoder, this effectively requires the contraction of a three-dimensional, rather than a two-dimensional, tensor network. Algorithms for such calculations have been developed in the context of condensed-matter physics, \cite{jiang_accurate_2008, pizorn_time_2011, orus_exploring_2012, xie_coarse-graining_2012, werner_positive_2016, kshetrimayum_simple_2017} and could potentially be applied here. 




\medskip
\noindent{\em Acknowledgements. --- } This work was supported by the Army Research Office Contract No. W911NF-14-C-0048.

\bibliography{physics}
\bibliographystyle{apsrev4-1}

\end{document}